# How to test cognitive theory with fMRI.

**Authors:**
Christopher H. Chatham[1] & David Badre[1,2]

**Affiliations:**
[1]Cognitive, Linguistic & Psychological Sciences
[2]Brown Institute for Brain Science
Brown University

*"If the mind happens in space at all, it happens somewhere north of the neck. What exactly turns on knowing how far north?"* (Fodor 1999)

*"… to say that neuroimaging answers only the 'where' questions is to confuse the superficial format of raw neuroimaging data with the content of the questions those data can answer; Neuroimagers collecting fMRI data need no more restrict themselves to "where" questions than cognitive psychologists measuring reaction times need limit themselves to "when" questions."* (Mather, Cacioppo & Kanwisher, 2013)

## Introduction

The objective of this chapter is to provide a guide to using functional magnetic resonance imaging (fMRI) to test hypotheses motivated by theories of cognition. This is, of course, a daunting task, as the premise itself – that fMRI data can inform cognitive theories – is still actively debated. Below, we touch on this debate as a means of framing our guide. In particular, we argue that the hypotheses motivated by theories of cognition can be constrained by neuroscientific data, including that offered by fMRI, but to do so requires embellishing the cognitive theory so that it can make predictions for neuroscience; much the same as how testing a cognitive theory using behavior requires embellishing that theory to make experimentally realizable behavioral predictions (i.e., the process of generating operational definitions). Moreover, recent years have seen the development of several new approaches that allow fMRI to better test neurally embellished models. Along with a review of several ways of testing neurally embellished cognitive theory using fMRI, we also consider the inferential challenges that can accompany these approaches. Readers of this chapter should gain an understanding of both of the potential power and the challenges associated with fMRI as a cognitive neuroscience methodology. As such, this guide should provide an informed basis for applying these approaches to the development of more powerful and explanatory theories of the mind.

## "What can fMRI tell us about the mind?"

Although the history of studying the brain to address psychological questions has its origins at least as far back as the field of psychology itself, it seems undeniable that the last two and a half decades have seen a massive expansion in the prevalence, scope, and impact of the cognitive neuroscience approach. Much of this growth is attributable to the advent of a single tool: fMRI.

The development of fMRI permitted psychologists to measure brain activity using largely the same types of tasks that they were using in behavioral settings and in typical healthy populations. Moreover, the method was accessible to a wider range of scientists than previous neuropsychological and neuroimaging approaches. Thus, the use of fMRI expanded rapidly, and has also expanded the scope of testable cognitive hypotheses along the way. More broadly, fMRI has changed how studies of cognition affect the public – both via its reporting in the media, and via its influences in public policy (e.g., the Mental Health Parity and Addiction Equity Act of 2008) and in the courtroom (Racine, Ilan & Illes, 2005; McCabe & Castel, 2008; Weisberg, Keil, Goodstein, Rawson & Gray, 2008; Beck, 2010; Aron et al., 2007; Roskies, Schweitzer & Saks 2013). Though not all of these impacts are universally positive, the impact is undeniable.

Along with these changes has come a substantial amount of academic controversy, some of it surrounding the challenge of whether fMRI has or can ever inform cognitive theory. Many question whether the impact of fMRI data on cognitive theory has been any greater than what could have been acquired from purely behavioral experiments, and indeed, whether such an impact is even possible in principle (Churchland & Sejnowski, 1988; Coltheart 1999; 2004; 2006; 2010; 2011; 2013; Cooper & Shallice, 2010; Bechtel, 2002; Levin & Aharon, 2011; Mole & Klein, 2010; Machery, 2012; Page, 2006; Tressoldi et al., 2012; Uttal, 2001; Satel & Lillenfeld, 2013). Some even suggest that use of fMRI is currently premature, and should not be pursued until a complete cognitive theory has been developed (Loosemore & Harley, 2010).

Skepticism regarding the impact of fMRI on cognitive theory generally comes in two forms. First, there are those who claim that cognition is most gainfully analyzed at a level more abstract than its neural implementation. By this view, which involves a strong assumption of multiple realizability (e.g., Marr, 1982), neuroscientific data only explain how cognitive processes are implemented, and cannot inform theories about the identities or interrelationships of cognitive processes. For this group of skeptics, fMRI is only the most prominent example of human neuroscience data. However, no other neuroscience methodology would fare much better in constraining cognitive theory. The second group of skeptics may concede that neuroscientific data could hypothetically constrain cognitive theory. However, they are unconvinced that the type of neuroscience data offered by fMRI, given its well-known limitations, is capable of providing such constraints.

This kind of debate is important, particularly given the resources and focus on fMRI in the human cognitive neuroscience literature. Nevertheless, we suspect that many versions of this debate may remain unresolved partly because the strongest forms of these challenges build in a certain circularity. Specifically, challenges to the idea that fMRI can inform cognitive theory generally assume as a starting point that cognitive theories can be fully described using terms that are wholly independent of any and all neural implementations. This premise, however, renders the argument circular: for granted, any complete theory that can be fully described without reference to the brain is by definition incapable of benefitting from brain data, because that theory manifestly makes no necessary claims regarding its neural implementation! To put it bluntly, such a purely cognitive theory would remain unaffected by neuroscience data even if it were shown that all of cognition occurred *south* of the neck – such as in the nerves of the bladder rather than the brain (see Buchsbaum & D'Esposito, 2008; Buchsbaum, 2012 for similar arguments). So, beyond expressing skepticism that such a complete theory of cognition, so defined, will ever be found, we will avoid this rigged debate and instead consider in what way neuroscientific data can be made useful to cognitive scientists.

To test a cognitive theory using any neuroscience method, including fMRI, one requires what Coltheart (2012) has termed a "neural embellishment". In other words, a theory must make a commitment to some testable neuroscientific correlate of the theory. To be clear, this embellishment is similar to the traditional concept of an operational definition, in that it specifies how some conceptual component of a model can be measured. Thus, just as a classical operational definition does not require a fully elaborated mechanistic model, a neural embellishment does not require a detailed neural implementation (though as we note later in the chapter, having such a model can be quite helpful).

For example, a model in which delayed discounting relies on interaction among separate systems – one driven by immediate rewards and the other that rationally computes value – might be quite naturally embellished to predict that distinct neural systems will correlate with each kind of valuation. Indeed, data from fMRI has provided some evidence along these lines (e.g., McClure et al., 2004). In this case, some degree of support is offered for the cognitive model simply by knowing that the brain handles these valuations differently – consistent with the theory's most direct and natural neural embellishment. Of course, further work is required to verify the specificity of the link, and to test further embellishments (e.g., regarding the contribution of various neurotransmitter systems, underlying neural representations, or even more specific features of the system). But, such systematic reduction in uncertainty is part of the normal progress of science, and is not unlike that experienced by the purely behavioral scientist.

**Why should one embellish a theory?**

These debates about the value added by fMRI to cognitive psychology also sidestep a key issue: why one might *want* to neurally embellish otherwise "purely cognitive" theories. As will be seen, the answers to this question are similar to why one might want to *behaviorally* embellish an otherwise purely abstract account of cognition (see also McClelland et al., 2010, and Meehl, 1978). While there is undeniably much value in purely cognitive theorizing, neural embellishment can both add inferential power and mitigate inferential risk. We now elaborate a partial list of reasons why a cognitive scientist might want to neurally embellish their cognitive theory.

*Neural embellishment mitigates the risk of descriptive abstractions.* The underlying assumption of a "purely cognitive" theorist is that the final product of a purely cognitive psychology – i.e., any formalized theory of cognition which suffices to explain all behavior – will be an explanation of the emergent phenomenon that the brain (or brain and body together) implements. But this is far from guaranteed, even if the formalized theory is accurate to an arbitrary degree. For example, Ptolemy's system of epicycles can explain planetary motion (or indeed *any* continuous and periodic motion; Hanson, 1960) to arbitrary accuracy[1], but it is still an utterly wrong theory of planetary motion in terms of the underlying physical dynamics. Similarly, the cognitive ontology entailed by classic cognitive psychological theories could amount to mere descriptive abstractions – i.e., mere approximations of the features that emerge from real underlying cognitive processes. The danger here is that these abstractions could lead research programs towards phenomena that do not have physical reality whatsoever, in the same way that Ptolemaic epicycles could have lead astronomers to search the sky for orbits that do not exist (McClelland et al., 2010).

*Out-of-sample tests can help resolve model mimicry.* Model mimicry – whereby dissimilar models can predict similar things for a given data set – may often thwart progress in some theoretical controversies (e.g., Greenwald, 2012). Model mimicry occurs because, in general, theory is underconstrained by data (Goel, 2005; Pylyshyn, 1979), and numerous theories can be adapted to account for any existing set of data (Figure 1). But by testing a theory beyond the sample domain it has already been designed to account for – in our case, by neurally embellishing an otherwise purely cognitive theory – one can distinguish models that might normally make similar predictions (e.g., for within-sample tests; see also White & Poldrack, 2013).

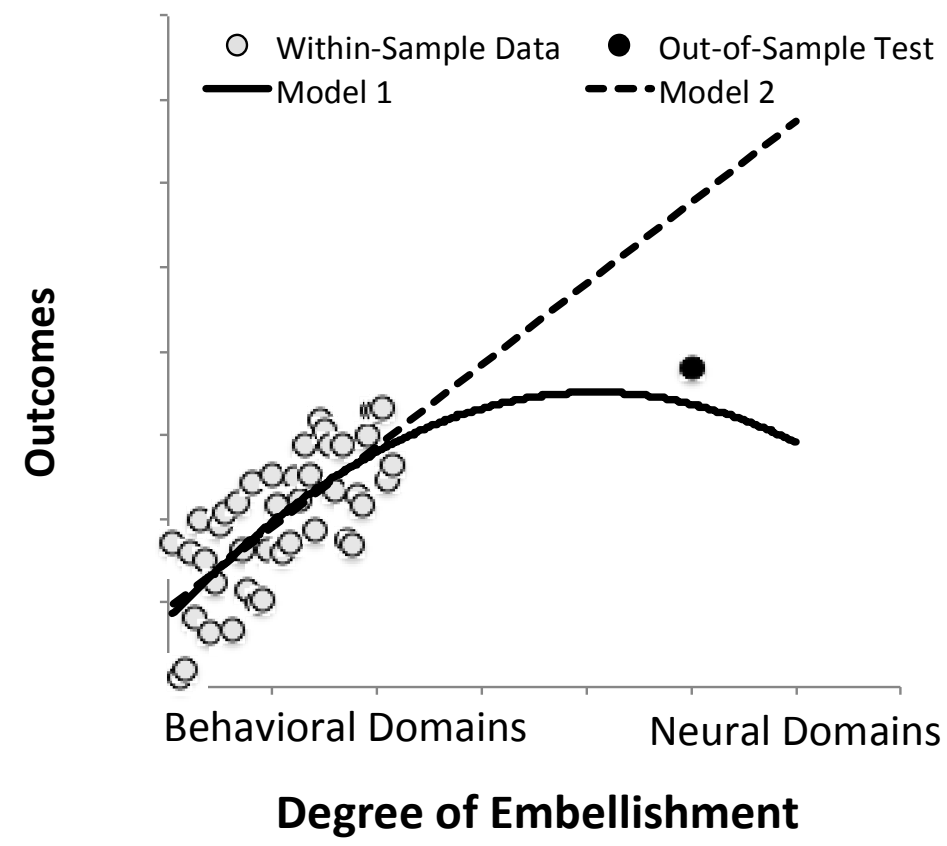

**Figure 1.** Model mimicry can thwart model comparison on the basis of existing data, but models can often be more productively compared with an out-of-sample test. Such tests require that at least one theoretical model can be embellished in whatever way is necessary to make this new domain relevant.

*Purely cognitive theories rest on non-falsifiable assumptions.* The justifying assumption of purely cognitive theories is a form of multiple realizability: that neural "hardware" can always support a viable cognitive account (e.g., at Marr's algorithmic or computational levels of description; 1982). But for this assumption to be categorically correct, the brain (or at least those parts involved in the cognitive process of interest) must be shown to be (at least) equivalent to a Universal Turing Machine (Henson, 2005; Eliasmith, 2003). In this way, pure cognitive theorizing relies on an assumption that the brain or its relevant parts are computationally universal – yet because this implementation is not specified, the assumption of its universality also cannot be falsified (see also Searle, 1990 and Bechtel & Mundale, 1999 for similar arguments).

*Multiple realizability does not imply infinite realizability.* Even if there are many possible implementations of a given process, there may not be an infinite number of them. For example, while a *purely* symbolic representation of mental imagery might conceivably be subserved by any of many different cortical areas, it is exceedingly unlikely to be subserved by primary visual cortex, which is known to encode visual features in a depictive manner. Thus, the necessity of primary visual cortex for mental imagery argues strongly against the notion that purely symbolic representations (as opposed to at least partially depictive) subserve mental imagery (Kosslyn et al., 1993). The broader point is that even if there are numerous viable implementations of a given cognitive theory, there may nonetheless be neural data that are informative by virtue of being *inconsistent* with that theory (Houng, 2012; Henson, 2005).

[1] A particularly compelling example of the risk of descriptive abstractions can be seen here: http://www.youtube.com/watch?v=QVuU2YCwHjw

*Integrative work succeeds.* The possibility that the mind can be independently explained at multiple levels of analysis is sometimes taken as a *recommendation* for scientific inquiry (Churchland & Sejnowki, 1998; Churchland, 1988; Colombo & Series, 2012; Haig, 2005) even though the history of science amply demonstrates that integrative work, spanning multiple levels of analysis, is a particularly productive agenda (see Bechtel, 2002 and Bechtel & Richardson, 2010 for similar arguments).

*Good theories usefully predict (and explain) testable experiments.* Many useful and testable predictions are implementational in nature – e.g., the effects of various drugs, or of aging. A very useful cognitive theory should explain those phenomena. This is not a qualitative departure from existing cognitive theories, because such theories have already been embellished to explain other implementational details of the system. This includes, to name just one general example, the influence of the human body's structure on the cognitive system (e.g., Rumelhart & Norman, 1982; Lohse, Jones, Healy & Sherwood, in press). Embellishing cognitive models to account for known biology not only makes use of this rich source of data, but also enables a model to make predictions regarding direct manipulations of the biology itself (e.g., via transcranial magnetic stimulation or targeted pharmacological manipulations) – manipulations that are crucial for the progression of both basic science and applied work alike.

*Neuroscience needs theory.* Neuroscience is accumulating a trove of data. But in the absence of neurally-embellished cognitive theories, this endeavor can sometimes resemble a kind of vacuous phenomenology, or "explanationless collections of observations — that is, mere 'stamp collecting'" (Ashton, 2013). Neuroscientific findings, including fMRI, obviously rely on cognitive theory to escape this fate (Wixted & Mickes, 2013).

*Non-embellished theories are at a consistent explanatory disadvantage.* To neurally embellish your theory is to enable it both to benefit from neuroscience data and to possibly be falsified by it – thereby allowing it to more strongly contribute to the progression of science. By contrast, any non-embellished theory is at an immediate disadvantage. To see why, consider the so-called "consistency fallacy" (Mole & Klein, 2010; Coltheart, 2013). This fallacy is putatively committed when one claims that some piece of evidence supports a theory without showing either how an alternative outcome could have contradicted that theory, or how the evidence contradicts another theory (Figure 2A). But from a Bayesian conception of evidence (Figure 2B), there is no such fallacy: our belief is rightly increased in a theory when an experiment's outcome is merely consistent with it (Figure 2C) – even when no outcome could possibly be inconsistent with any other theory under consideration. Conversely, our belief is also rightly decreased in that theory to the extent the outcome is posited to be unlikely – even if it falls short of contradicting the theory (Figure 2D).

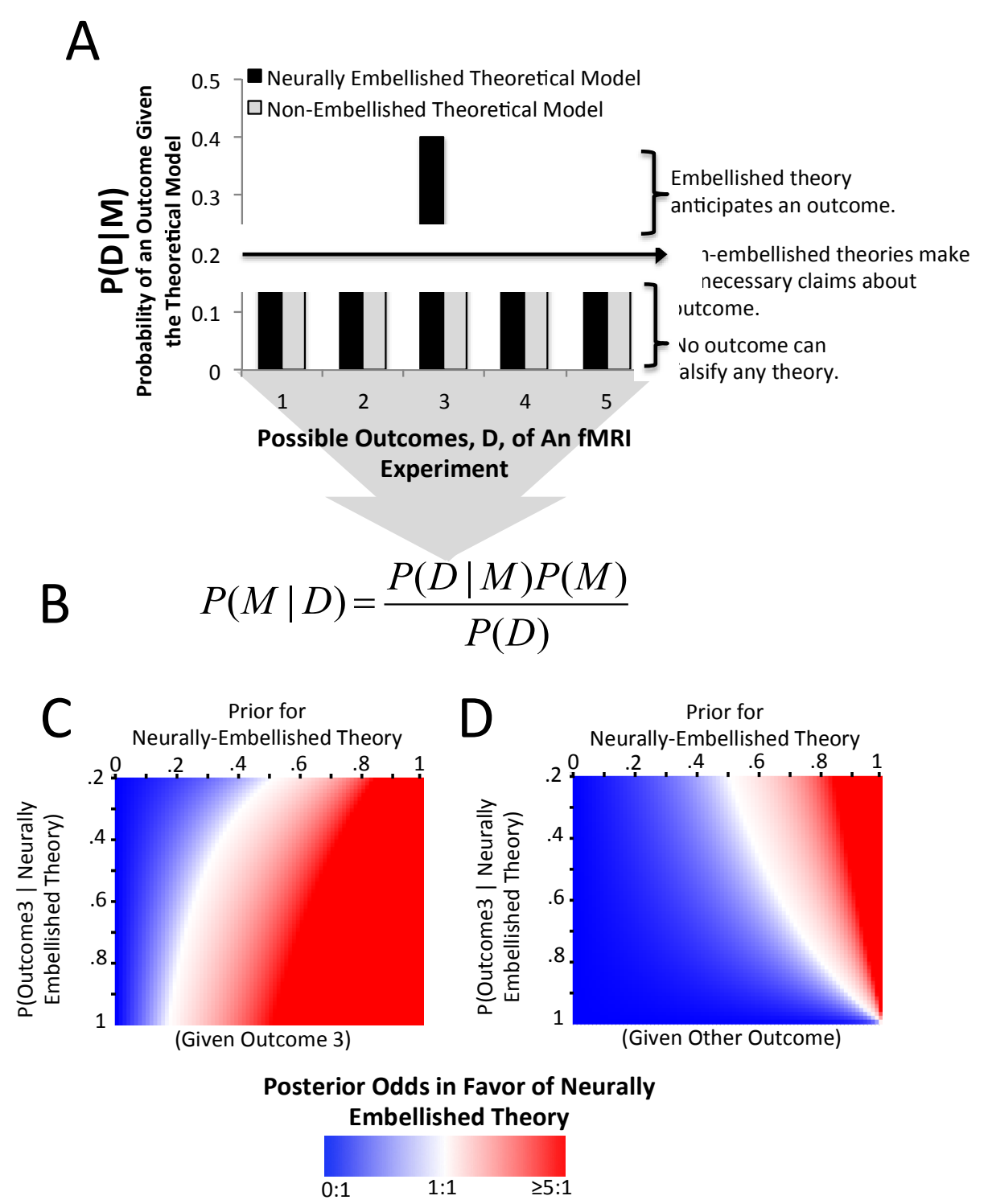

**Figure 2. It is no fallacy to claim that merely consistent evidence supports a theory. A.** One theory is non-embellished, and thus makes no differential claims about any outcome of an fMRI experiment. A second theory has been weakly neurally-embellished – it assigns slightly higher likelihood to some outcome. **B.** Bayes' rule tells us how to optimally change our belief in each theory, depending on the outcome of our experiment (P(D)), our prior belief that each theoretical model was the correct one (P(M)), and the probability of the outcome given that theoretical model (P(D|M)). **C.** If the experiment yields an outcome consistent with the most likely prediction of the neurally-embellished model, it is generally more likely that model is correct – even though no possible outcome would have falsified any model under consideration. **D.** This cuts both ways, however; our relative belief in the non-embellished theory is somewhat greater when the evidence is more consistent with the observed outcome. Thus, perhaps counterintuitively fMRI data *can* inform even non-embellished theories - so long as there is some valid neural embellishment of an alternative theoretical model.

To make this concrete, if a cognitive theory of working memory were neurally-embellished to suggest that recruitment of prefrontal cortex was more likely than not during working memory, the clear involvement of prefrontal cortex in working memory (e.g., Braver et al., 1997) should rightly increase our belief in that theory – even though a failure to observe prefrontal recruitment would not have falsified it, and even though non-embellished theories make no necessary claims about the brain whatsoever. Put succinctly, Bayesian reasoning makes formal the dictum that "vague theories can be only vaguely right" (Gallistel, 2009) because "the most restrictive consistent hypothesis is strongly favored" (Perfors, Tenenbaum, Griffiths & Xu, 2011).

*Neuroscience can directly test the key claims of cognitive psychology: whether (and which) internal states intervene between stimulus and response.* A cornerstone of the "cognitive revolution" was that internal representations and transformations can be scientifically studied even if only stimulus and response are directly measurable. Cognitive theories about these intervening states were thus behaviorally embellished, by necessity. But if a cognitive theory is not expressed in terms that can also be tested neurally, *then there is no way to observe these intervening states in isolation.* In other words, we can examine these latent states only as they exist within a cascade of other possible intervening states that lead to a response – a situation where classic additive factors logic may be readily defied, and assumptions of pure insertion violated (McClelland, 1979; an assumption not intrinsic to some fMRI designs; Price & Friston, 1997; Badre, 2011). The simplest example of these advantages is that fMRI can reliably characterize internal representations even when no overt response is being produced (e.g., during passive mental imagery; Reddy, Tsuchiya & Serre, 2010). However, to interpret this neural response with respect to cognitive theory requires a neural embellishment. In other words, a commitment to neural implementation is required that allows one to test "responses" of whatever intermediate states are posited by the cognitive theory, without a potentially confounding influence from other processes. Hence, lacking a ready means of neural embellishment ultimately limits the opportunities available to develop and compellingly test a cognitive theory (see also Mather, Cacioppo & Kanwisher, 2013).

We wish to emphasize again that important and fundamental progress can be made in cognitive science without a neural embellishment. However, we also think that, for the reasons given above, neurally embellishing one's cognitive theory confers a number of advantages that should make doing so attractive to a cognitive scientist. Thus, we now consider how one might go about embellishing a cognitive theory.

## Ways to Neurally Embellish a Cognitive Theory

Theories are intrinsically underconstrained by data (Greenwald, 2012; Goel, 2005; Pylyshyn, 1979), so any embellishment is useful if it allows more data to speak to a theory. But how is neural embellishment to be accomplished?

There are two levels to this general problem. The first is how to express a cognitive theory in a way that makes some commitment to a neural implementation. This part of the problem is a matter of much philosophical debate, with some arguing that there is no general set of laws that bridge these levels of analysis (see McCauley, 1998 for a summary), and others who posit such principles are discoverable (Bressler & Tognoli, 2006; Atallah, Frank & O'Reilly, 2004). The more specific problem considered here is how cognitive theory can be neurally embellished in a way that specifically enables an effective test with fMRI. The (arguably) smaller impact of fMRI on theory in cognitive psychology, as compared to its practice, could imply this part of the problem is perniciously difficult.

A rough guide to neural embellishment is in fact provided by the behavioral embellishments of otherwise abstract cognitive theories. Generally, some experimental manipulation is predicted to influence the demands on a particular cognitive process (Figure 3A). The demands placed on this process will, by virtue of that process's position in some theorized cognitive architecture, exert an influence on behavior (Figure 3B). Both kinds of influences – the influence of a task manipulation on a cognitive process, and of that cognitive process on behavior – are often assumed to be linear with a varying slope and intercept, but could in principle be either more constrained (e.g., the identity function y=x) or more flexible (e.g., nonlinear, even non-monotonic).

Both classical cognitive psychological theories as well as more recent Bayes-inspired cognitive theories can be understood within this very general framework. For a classical cognitive example, consider Egeth & Dagenbach (1991), who showed that degradations in the visual quality of nontarget items produces linear costs to target identification processes, but that these processes can operate in parallel – consequently yielding a nonlinear (subadditive) effect on behavior. For a more recent Bayesian example, consider the work of Gershman & Niv (2012; Exp 3), in which rewarding experiences are posited to exert a non-linear influence on internal reward predictions (where the functional form of this influence is determined, in part, by Bayes' rule), and these reward predictions are in turn posited to exert a non-linear (sigmoidal) influence on subsequent choices.

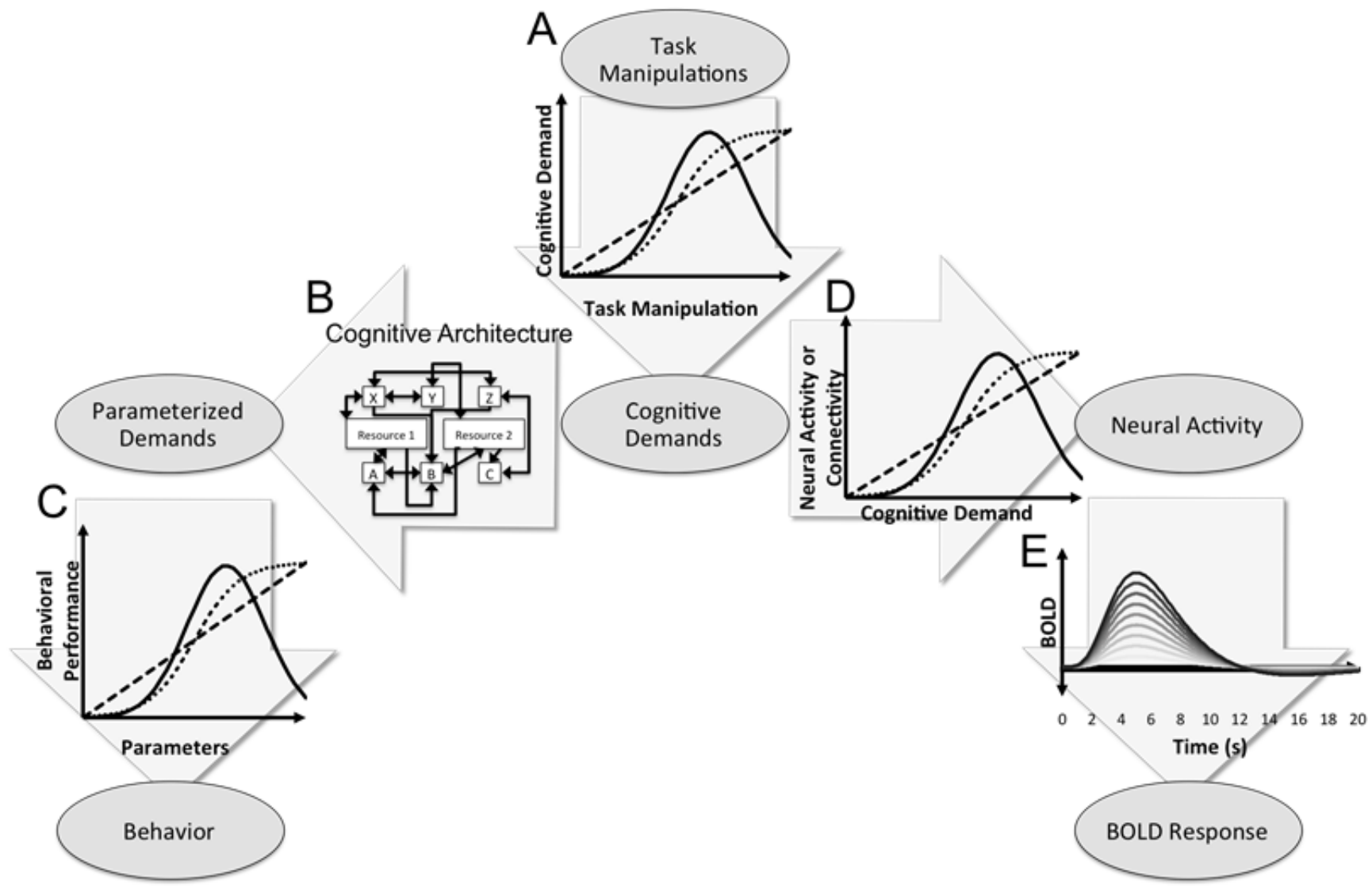

**Figure 3. Cognitive theories require embellishment to be tested. A.** Task manipulations are assumed to induce some kind of cognitive demand, though this function may not be linear or even monotonic. **B.** By virtue of a theoretical cognitive model, this demand may influence multiple cognitive processes. **C.** These influences have some kind of effect on behavior, whether linear or not. **D.** One straightforward neural embellishment of a cognitive theory is to specify the relationship of cognitive demand to perisynaptic neural activity in a region, or correlated changes in neural activity across regions. **E.** The subsequent mapping to the measured BOLD response is well approximated by a linear time invariant system of equations (such as the canonical double gamma hemodynamic response function, illustrated here).

Like these kinds of behavioral embellishments, neural embellishment requires specifying a series of functional relationships. Most commonly, this is simply assumed to be a linear function relating one kind of cognitive demand to perisynaptic neural activity (not necessarily neural firing rate; e.g., Logothetis, 2008; see also Ekstrom, 2010) in a particular region (Figure 3C). Note that this “single-association logic” need not assume that the cognitive process of interest is the *only* cognitive process that influences neural activity in this region (e.g., Henson, 2005).

Alternatively, neural embellishment may specify a functional relationship between cognitive demands and a more complex neural entity. For example, cognitive demands might be posited to relate to an axis in neural space – say, the location of peaks in perisynaptic neural activity along the rostral-to-caudal axis of the prefrontal cortex (e.g., Badre & D’Esposito, 2007). One can also specify how cognitive demand relates to some characteristic of neural activity in a region other than its mean – for example its shape (Friston, Josephs, Rees & Turner, 1998; Glover, 1999; Woolrich, Behrens & Smith, 2004; Lindquist & Wager, 2007), its variability (e.g., Dinstein et al., 2012; Mueller, et al., 2013; Chatham, Frank & Badre, under review; Garrett et al., 2013), its correlation with perisynaptic neural activity in other regions (e.g, Kahnt, Park, Burke & Tobler, 2012; Chatham & Badre, 2013), the information decodable either from the region (e.g., Tong & Pratte, 2012; Chatham et al., 2012; Öztekin & Badre, 2011) or from its correlations with perisynaptic neural activity in other regions (Coutanche & Thompson-Schill, 2013), or the relationship of any of these factors to behavior or one another.

Even an entire cognitive architecture's topology or connectedness can be assessed if it is neurally-embellished and the appropriate graph theoretic (e.g. Bullmore & Sporns, 2009) or representational similarity metrics (Kriegeskorte & Kievit, 2013) are used. Moreover, all of these techniques are generally possible using standard fMRI acquisition parameters. Together, they demonstrate the wide range of opportunities for testing neurally-embellished cognitive theories with fMRI.

The remaining functional relationship – from perisynaptic neural activity to the BOLD response actually measured by fMRI – is comparatively simpler. Though there are some known exceptions (e.g., Birn et al., 2001; Huettel and McCarthy, 2000; Miller et al., 2001; Vazquez and Noll, 1998; Wager et al., 2005), this mapping is well-approximated by a linear time invariant system: predicted perisynaptic neural activity can be straightforwardly convolved with a hemodynamic response function (Figure 3D; see e.g., Boynton, Engel, Glover & Heeger, 1996; Logothetis, 2008). Thus, the difficulty in neurally embellishing a cognitive theory is largely in specifying the mapping between a cognitive resource and the spatiotemporal structure of neural activity. As such, the core challenge is not distinct in kind from that posed by embellishing a cognitive theory to account for spatiotemporal aspects of behavior – say, reaction times across a subjects' arms, hands, or fingers during the course of a task.

## Testing Neurally-embellished Cognitive Theory with fMRI

How can we be sure our neural embellishments are valid? Put another way, how do we know when a cognitive theory is invalidated by the results of an fMRI experiment, as opposed to merely its neural embellishment? It is instructive to note that – again – formally similar issues confront classical cognitive theories with behavioral embellishment. For example, a large class of mean reaction time patterns predicted by parallel models of cognition can be mimicked by appropriately-constructed serial models (Townsend & Wenger, 2004; White & Poldrack, 2013). More generally, a comprehensive review of 13 famous theoretical controversies across the past 40 years of psychological research suggests similarly dismal prospects for other debates, *except* where new methods can be brought to bear on existing theories (Greenwald, 2012) – perhaps because they act as a kind of out-of-sample test (see Figure 2, above). Neither neural nor behavioral embellishments are immune to this basic form of inferential threat, though the specific kinds of inferential threats may differ, as we describe below.

To this end, here we provide a few approaches to testing cognitive theory along with select examples of how fMRI has informed (neurally-embellished) cognitive theories for each. We hope these examples will serve to illustrate many techniques in this enterprise: key ways that cognitive theory can be neurally embellished, how these embellishments can be inferentially tested, and the key assumptions that underlie these inferences – along with their boundary conditions. We highlight these inferential risks not to detract from the methods or discourage their use. No approach in all of the cognitive or neural sciences is immune from inferential limits. Rather, we view each technique as a powerful means of enabling future developments in cognitive theorizing, and understanding the strength of evidence they provide is essential to applying them within a research program.

*Univariate BOLD Dissociation Logic.*

A foundational question for cognitive theorists in any domain is *how many* latent states may intervene between stimulus and response for a given psychological phenomenon. Dissociation logic is one approach to answering such questions of dimensionality. The "logic" itself is deceptively simple: if two experimental manipulations have dissociable influences on a dependent measure, then the dependent measure must reflect the combination of (at least) two processes. By contrast, in the absence of such dissociable effects, then a unitary model of the underlying states cannot be rejected. Many patterns of fMRI data can imply an underlying dissociation, though some do so more strongly than others; this ranking of dissociation evidence is discussed more extensively elsewhere (Dunn & Kirsner, 1988; Henson, 2005; Machery, 2012). However, at one extreme of the ranking, a single dissociation is considered the weakest evidence for independence. A single dissociation in fMRI, wherein an experimental manipulation affects activation in one brain region more than another brain region, can readily be accounted for by a single process model in which both brain areas respond to the same underlying process, but differ in their responsivity (for a similar argument as it applies to behavioral data, see Wagenmakers, Krypotos, Criss & Iverson, 2012). At the other extreme, a cross-over double dissociation is considered the strongest evidence of independence. A double dissociation in fMRI, wherein one brain region activates more for manipulation A than B and a second brain regions activates more for B than A, is more difficult to account for with only a single process (discussed further below). Dissociation logic has the added advantage of being tested using the most classic type of fMRI analysis: the simple univariate (mean) BOLD contrast between conditions.

The domain of cognitive control is one clear example of fMRI's utility in questions amenable to dissociation logic. For example, it remains controversial whether the human capacity for planning and abstract problem solving relies on a componential architecture of executive functions that support cognitive control (e.g., Miyake et al, 2000) or a unitary system. Recently, it has been proposed that two key components underlie difficulties with planning: the degree to which the relative ordering of actions can be deduced directly from the desired goal (termed "goal hierarchy ambiguity"), and the degree to which these actions may require intermediate moves with interdependencies ("search depth;" Kaller, Rahm, Spreer, Weiller & Unterrainer, 2011).

The distinction between these concepts can be clarified by example (here adapted from Kaller et al., 2013). Suppose you are planning a three-course meal for friends. The problem of *serving* the courses has relatively little goal hierarchy ambiguity, because appetizers must come before the main course, and the main course before desert. Greater goal hierarchy ambiguity is involved in the problem of serving the side-dishes for the main course: the goal state of serving a three-course meal does not directly specify this ordering. By contrast, search depth is better illustrated by the cognitive process required during *preparation* of this elaborate meal. If the courses and their side-dishes require overlapping sets of kitchen appliances, then one needs also to consider intermediate subgoals like washing these appliances before preparing the next side-dish or course. This would involve potentially large search depth conducted at the subgoal level, to ensure the meal was prepared efficiently. .

Reaction times in a task designed to dissociate search depth and goal hierarchy ambiguity (specifically, a version of the classic Tower of London task) yield roughly additive effects of these two factors, but that result is equally consistent with independent cognitive resources for depth and ambiguity as it is with the idea that both factors require a single underlying resource due to "planning difficulty". However, an fMRI experiment revealed a cross-over double dissociation of these processes (Kaller, et al. 2011). Whereas the left DLPFC showed a larger BOLD response as a function of goal hierarchy ambiguity than search depth, the right DLPFC showed the opposite effect.

These findings are transparently more consistent with theories that posit a distinction between these subprocesses than with those that posit no such distinction. The effect was regionally specific – it was not observed in the nearby frontal eye fields – further affirming this rather specific neural embellishment. Finally, the inference of distinct subprocesses in planning was principally enabled by fMRI, given that the behavioral data were ambivalent on this point.

*Dissociation Logic – Inferential Risks.* Kaller et al. (2011)'s results are something of a marvel. Such qualitative double dissociations are rarely observed empirically; even subtle departures from this ideal pattern, including those that nonetheless still show crossover interactions, render the results susceptible to a variety of single process interpretations. As noted previously, this has lead to several rankings of the strength of evidence of independence provided by various dissociations (Dunn & Kirsner, 1998; Machery, 2012), though these rankings have not gone without criticism (McCloskey, 2001; Davies, 2010).

Unfortunately, even the ideal cross-over double dissociations, like Kaller et al's, leave room for uncertainty. A thought experiment makes this clear. Let us imagine a world where both putative subprocesses in fact produce demands on a single underlying process – say, the number of moves a subject must look ahead in order to visualize the goal. We will also assume that left DLPFC is preferentially recruited when the number of look-ahead moves is small, and the right DLPFC is preferentially recruited when the number of look-ahead moves is large. If we further assume that the goal hierarchy ambiguity manipulation increases the number of look-ahead moves only by a small amount, and the search depth condition increases the look-ahead moves by a large amount, then these conditions can be understood as ordered along a single "look-ahead" axis (Figure 4A). When neural activity is taken from each region at these points along the axis, it can yield precisely the kind of "pure" double dissociation described by Kaller et al (Figure 4B).

The point of this thought experiment is not to seriously propose that left and right DLPFC are preferentially recruited at different points along a "look-ahead" axis. Rather, the point is that there will always be alternative single-process explanations: in principle, any "pure" double dissociation can be explained in this manner. Ruling this out requires more data.

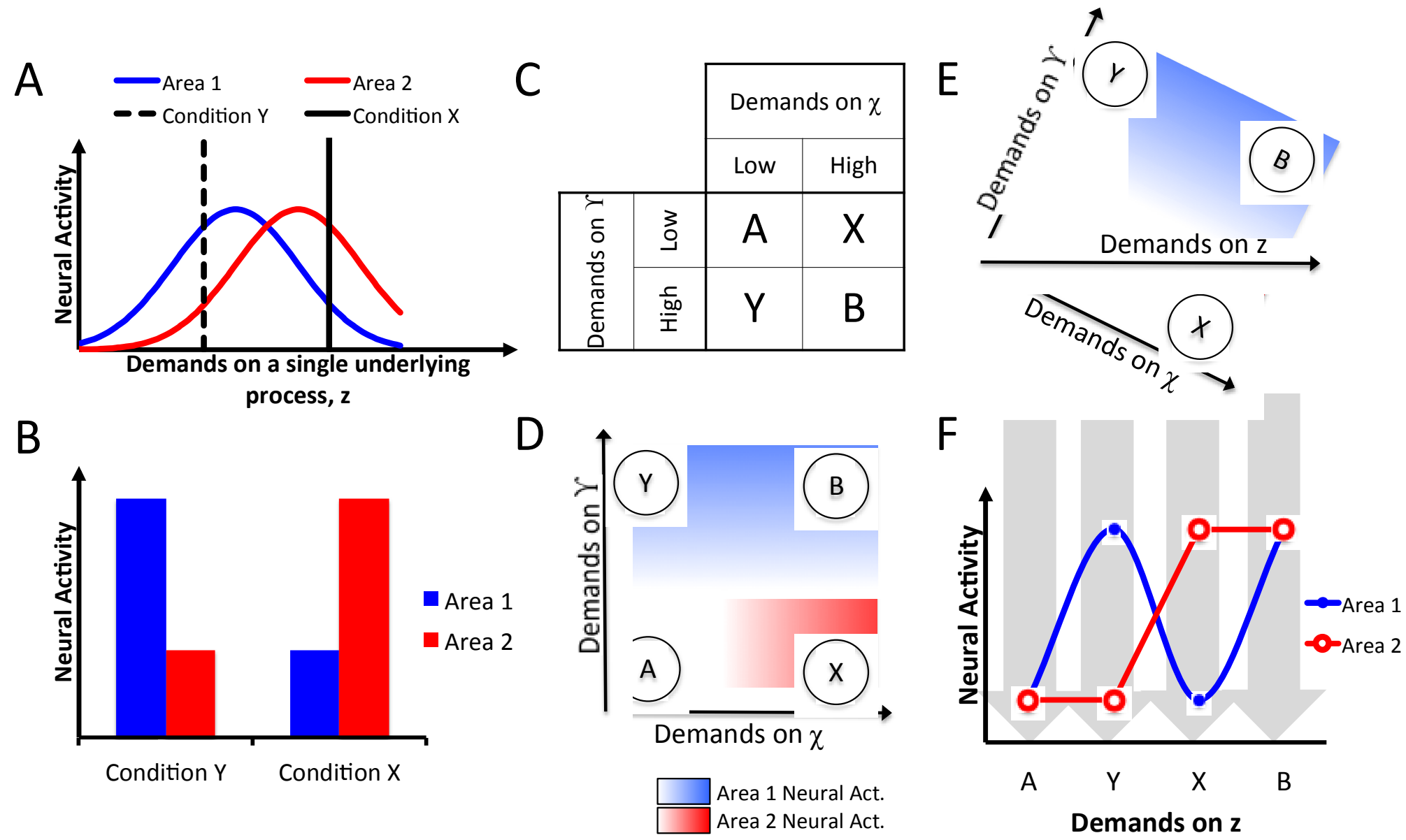

**Figure 4. Single-process accounts of a pure 2x2 double dissociation can in theory be ruled out with a factorial 2x2x2 design. A.** Two conditions can be ordered along an axis corresponding to demands on a single underlying process, and the neural activity induced in two regions assumed to be non-overlapping and non-monotonic functions of those demands. **B.** These assumptions suffice to explain pure double dissociations across two regions in two conditions. **C.** Under a canonical two-process account, each condition places demands on separable cognitive processes; these can be factorially manipulated in a 2x2 design. **D.** A two-process account predicts completely orthogonal mappings from demands on these processes to neural activity in the two regions. **E.** Single-process accounts amount to assuming that there is a rotation of the design matrix from which these conditions can be projected onto a single axis without a meaningful loss of information, much like the rotations used in extracting a principal component from several indicators. **F.** Any rotation will require neural activity to follow (at least) a cubic function of demands on that underlying process, in one of the two areas, if the results expected from a two-process account are observed. It is seems unlikely that such a pattern could ever be adequately justified theoretically by a single-process account.

A factorial design is actually sufficient for this purpose. Thus, what was a 2 (condition) x 2 (region) interaction can be expanded into a 2x2x2 design (Figure 4C). In such a design, the predictions of a two-process account are straightforward: manipulations of either cognitive process should not affect the involvement of the area which is putatively insensitive to that process (Figure 4D). Conversely, any claim that these 4 conditions can still be ordered along a single axis reduces to the claim that the design matrix can be rotated so that each condition projects to a distinct point along that new axis. In other words, a single-process account posits that two putatively distinct processes (i.e., the orthogonal axes in the design matrix) actually have varying demands on a single process (i.e., another axis that "cross-cuts" the design matrix under some rotation; Figure 4E). If the results expected from a two-process account are observed, a continued single-process interpretation must assume that recruitment of one of the two areas reverses *twice* as a function of demands on that single underlying process (Figure 4F) – an assumption of non-monotonicity so unlikely it renders a single-process explanation practically untenable. Formal methods for capturing these non-monotonic patterns in real data are a rapidly developing approach in dissociation logic (e.g., state-trace analysis; Staresina, Fell, Dunn, Axmacher & Henson, 2013).

In summary, dissociation logic is one of the most powerful, and indeed most common, approaches to testing hypotheses of dimensionality in a cognitive theory. This type of evidence is further strengthened by the use of fuller factorial designs – alternative single-process accounts must invoke increasingly complex forms to account for such data.

*Parametric Designs – An Example from Model-Based fMRI.*

Factorial designs are statistically powerful: they concentrate data collection to extreme values of a manipulation, and thereby minimize the Type II error rate. Yet they also entail a sparse pattern of sampling in the space of possible manipulations, which means the sampled data points will underdetermine the class of functions that may explain the data (e.g., Figure 4). As above, additional confidence is granted only by the collection of additional data. One means of collecting additional data is to assess experimental manipulations at multiple continuously-valued intensities – that is, to sample the data at many intermediate values of some independent variable. These *parametric* designs reduce statistical power, because fewer data points are collected at the manipulation's extremes, but also mitigate inferential risk: the increased number of samples more strongly constrains the family of curves that may explain them.

Parametric designs have been of particularly widespread utility in fMRI, where non-monotonic functions are sometimes argued to be especially plausible (Machery, 2012), and where cognitive models can readily be tested if they make specific predictions about a quantity that should be important for the cognitive system. Parametric designs used in this latter mode are sometimes called "model-based fMRI," because they represent a tight integration of quantitative cognitive models and neural embellishment. For example, trial-by-trial variations in some (putatively) cognitively-encoded value can be tested for a relationship to the BOLD response, in the same way that trial-by-trial variations in some experimental manipulation might also be tested for a relationship to RT.

A particularly strong use of this approach can be found in recent work on reinforcement learning . One class of reinforcement theories (e.g., Daw, Niv & Dayan, 2005) posits that, even in the absence of reinforcement or any overt response, an agent can construct a model of how the various states of the world are interconnected, and that this model can be used to shape predictions about the optimal response to take in response to a future stimulus. Unfortunately, such a theory is very hard to test using behavioral measures, because its distinguishing predictions occur at a time when no overt responses are made.

Hence, behavioral data cannot be obviously brought to bear on the crucial predictions of this cognitive theory.

This issue is exemplified by a study (Gläscher, Daw, Dayan & O'Doherty, 2010) in which subjects passively observed a series of fractal images following a probabilistic grammar. By the proposed cognitive theory, subjects were even then constructing a mental model of these transitions by minimizing so-called “state prediction errors” – that is, by minimizing the difference between the expected and observed image at each moment in time. But because no overt response is made at this time, there is no opportunity to directly test the underlying hypothesis.

Subsequently, subjects were told a few specific images would be particularly rewarding, and allowed to make binary choices among these images to maximize the likelihood a rewarding image would appear. As predicted by this account, even subjects' very first choices were not random, and were consistent with some degree of learning of the grammar. Of course, behavioral results at this point are incapable of distinguishing between the detailed proposed computations and any number of possible alternatives (including, for example, an episodic memory-mediated reconstruction of prior experiences at the time a choice was required).

In this case, fMRI data allowed a crucial test of the cognitive theory. Here, the neural embellishment is quite straightforward: the assumption is that if cognition avails itself of the computations prescribed by the cognitive theory, there should be a neural correlate of these prediction errors in the brain. That is, no neural region was specified, rather a whole brain search was conducted instead. In addition, the functional form relating these calculated prediction errors to neural activity was, straightforwardly, taken to be linear. Despite these somewhat simplistic and sparse assumptions, the data were nonetheless resoundingly consistent with the proposed theory. Homologous areas across both hemispheres of the lateral prefrontal cortex (specifically, within the inferior frontal junction) were found to robustly correlate with these prediction errors. These effects were observed even in the absence of any reward or response. Again, these are precisely the kind of data unobtainable from traditional behavioral measures, and moreover precisely the kind of data necessary to test the specific predictions of this kind of cognitive theory.

*Parametric Designs – Inferential Risks.* Like multi-factorial designs, parametric designs are another means of mitigating the risks inherent to basic dissociation logic. They have been particularly useful in testing models that make quantitative predictions about changes in latent cognitive variables over time (e.g., Daw, Niv & Dayan, 2005, as noted above). But they come with their own inferential risks, and not all parametric effects are equally informative, or as informative as the exemplary work of Gläscher described above.

Specifically, a parametric design is effectively calculating the slope that relates a (multi-valued, and often continuously-valued) experimental manipulation with the continuously-valued BOLD response. As famously demonstrated by Anscombe (1973), such linear slope estimates can be identical across many different underlying patterns of data, some of which may not correspond to a true “linear” effect whatsoever (see Figure 5A).

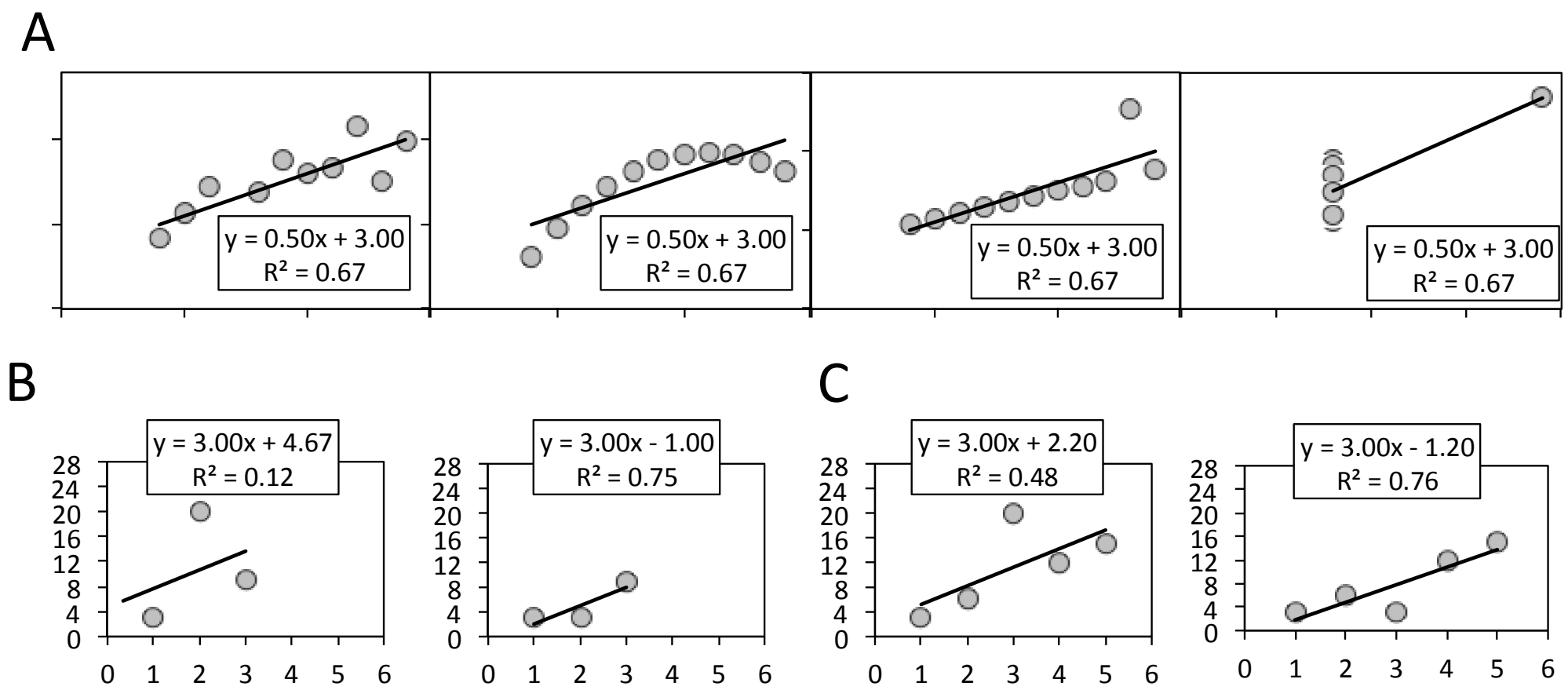

**Figure 5. Inferential risks with parametric predictors. A.** Anscombe’s quartet (1973) demonstrates how a variety of continuously-valued data points may lead to the same estimates of slope, intercept, correlation, as well as mean and variance. **B & C.** Discretely-valued and equal-interval predictors can be useful, but if there are an odd number of values to such predictors, the middle predictor will have no influence on least-squares slope estimates. By contrast, changes in the value of the middle predictor will always influence the intercept, and will generally influence the correlation estimate.

A related inferential risk is the case when a parametric predictor has only three evenly-spaced levels, or indeed any odd number of evenly-spaced levels. In such cases, the middle level of the predictor has zero influence on the estimated slope, and thus fully one-third of the data is ignored in its estimate. (By contrast, the correlation and intercept will generally be affected by the middle level of the predictor, as illustrated in Figure 5B&C). Thus, a linear effect alone may actually tell one nothing about whether the effect is truly parametric in the case of an odd-number of levels to a predictor. This may be particularly problematic when only a small number of levels to the predictor are tested.

The analysis of prediction errors (PEs) is generally robust to this particular problem (because they are continuously-valued). However, analysis of PEs illustrates another potential issue with parametric predictors which is particularly pernicious given the spectral properties of the BOLD response. Specifically, the distribution of PEs across the course of an experimental session is often highly skewed, with much larger PEs towards the beginning of the experiment (when expectations are perhaps most incorrect, relative to the true value), and much smaller PEs towards the end of the experiment (where subjects’ expectations may have converged closer to the true value). This poses a problem: such slow changes tend to fall within frequencies that are usually filtered-out by standard fMRI preprocessing pipelines, so as to eliminate the particularly strong sources of noise in the MR signal that occupy those low frequencies (Zarahn, Aguirre & D’Esposito, 1997). Conversely, the predominantly small and high-frequency PEs that occur when subjects’ expectations have approximately converged may actually fall *above* the frequency range over which BOLD can be detected with maximum sensitivity.

Thus, the spectral properties of MRI and BOLD can interfere with the analysis of some parametric regressors, particularly PEs, in multiple ways.

To minimize this issue some psychologists change the value of the predicted entities over the course of an experiment (a "nonstationary" design) but these designs may introduce new problems. For example, BOLD responses due to PEs and outcomes are often too highly correlated to distinguish, raising the possibility that some model-based fMRI results previously ascribed to PEs could actually reflect experienced outcomes (Erdeniz, Rohe, Done & Seidler, 2013; Wilson & Niv, 2015). Because non-stationary designs attempt to maintain a constant magnitude of PEs across a task, they will actually tend to reduce some of what little variance can be used to distinguish PEs from outcomes. As a result, nonstationary designs will tend to increase the collinearity of these terms, and thereby exacerbate the inferential ambiguity between PEs and outcomes. That said, in some cases this tradeoff may be acceptable.

Finally, recent work has highlighted an unexpected feature of model-based fMRI regressors: the neural correlates of these regressors can be surprisingly insensitive to variations in the learning rate parameter of the underlying reinforcement learning model (Wilson & Niv, 2015). In particular, the *minimum* correlation between the correct ground-truth PE regressor and one that could be derived from the same model with an incorrect learning rate is R=.7. On the one hand, this robustness mitigates some of the inferential risks associated with model-based fMRI, because the neural correlates of a particular regressor are unlikely to be drastically influenced by deviations from the best-fitting parameters, which may in turn be underdetermined by available data. On the other hand, this finding underscores the difficulties in distinguishing the neural correlates of PEs and of outcomes. For example, as noted by Wilson & Niv (2015), PEs are perfectly correlated with outcomes when the learning rate is zero and the reward distribution is fixed in an experiment. Thus, care must be taken in ensuring a given experimental design is capable of distinguishing the parameters of interest for a given model.

*Parametric Designs – Summary.* Parametric designs are a natural means by which to assess the most straightforward neural embellishments of well-specified cognitive theories, such as: do the quantitative components of a formal theory have direct neural correlates? A close fit to neural data might increase one's belief in a theory, but there are also clear inferential risks to this simple reasoning. Anscombe's quartet and related issues in the calculation of slope demonstrate that similar parameter estimates might emerge from a variety of underlying patterns, not all of which would constitute clear evidence for the neurally-embellished theory. To address these issues we recommend the use of parametric designs optimized to yield effects falling within the spectral range of BOLD's maximum sensitivity; parametric predictors that are either real-valued or have an even number of equal-intervals; and the use of non-stationary designs, with the caveat that extra care must sometimes be taken to evaluate collinearity across regressors. When carried out in this way, parametric fMRI designs have great potential for determining whether and how task-dependent cognitive constructs may be neurally encoded.

*Connectivity Analyses.*

So far we have discussed approaches to fMRI analysis that are merely "scaled-up" versions of the analyses often run on reaction time. Just as we assess the mean RT in two conditions, or the parametric fit of RT across these conditions to a model's predictions, we might assess the mean BOLD response or correlation of BOLD response with a parametric regressor, repeatedly so, across every voxel within the brain. Such analyses are sometimes termed "massively univariate."

However, one key advantage to fMRI is the enormous richness of the acquired data. This richness enables analyses with no clear or universal counterpart in the behavioral domain. For example, one can assess the degree to which the BOLD response in one region is correlated with the BOLD response in all other regions, and how that correlation may differ as a function of performance, experimental manipulations, their combination, or a host of other factors. Such analyses of "connectivity" are widespread, and of potentially great utility for testing the neural embellishments of non-modular cognitive theories. For example, modular accounts of cognition might be less easily embellished to predict or explain connectivity results than connectionism-inspired theories, where cognitive demands are more explicitly linked to differences in the correlations across neural substrates.

There are many applications of these connectivity approaches, some of them to domains where classical approaches to cognitive theorizing could never have been assessed (e.g., the "resting state," in which subjects experience no overt stimulus nor execute any overt response; see also Keilholz, Thompson, Magnuson & Pan, this volume). But a particularly successful application of connectivity to cognitive theory comes from the domain of working memory.

Cognitive theories of working memory sometimes highlight its paradoxical capacities: rapid, flexible updating in certain contexts, and robust, stable maintenance in others (e.g., Goschke, 2003). The incompatibility of these demands could imply that a computational division of labor supports them (e.g., Hochreiter & Schmidhuber; 1997), with distinct cognitive processes for updating and maintenance. In turn, a straightforward neural embellishment of this hypothesis might predict a corresponding *neural* division of labor; for example, that working memory updating involves a distinct kind of interaction across some set of neural regions than pure maintenance (e.g., Frank, Loughry & O'Reilly, 2001). Finally, if the theory is neurally-embellished to predict that working memory maintenance will involve the prefrontal cortex, then this theory can be readily tested using an fMRI technique known as psychophysiological interaction (PPI; O'Reilly, Woolrich, Behrens, Smith & Johansen-Berg, 2012).

In practice PPI amounts to a fusion between parametric designs (as described above) and functional connectivity (to be described below). The key idea is to predict the BOLD response from an interaction of the neural activity in a particular seed region and some experimental manipulation. The inference enabled by PPI is whether differences in the BOLD correlations between regions may differ across the various conditions of a task, or values of a parametric effect. In this particular example, a PPI analysis could reveal whether working memory updating is associated with greater coupling across regions than simple working memory maintenance, as expected by this neurally-embellished theory.

This study was recently conducted by Nee & Brown (2012). Consistent with the neurally-embellished theory described above (which is in fact better characterized as an emergentist or cognitively embellished neural theory [see below]; Frank, Loughry & O'Reilly, 2001), this study showed increased coupling between the prefrontal cortex and basal ganglia during working memory updating, relative to simpler maintenance. Unexpectedly, this was only the case when the to-be-updated information was particularly abstract (i.e., only very indirectly relevant for responses). By contrast, when updating of working memory with relatively more concrete information (i.e., information that could more directly specify a response), prefrontal and parietal regions increased their coupling instead. Together the results constituted a near-ideal double dissociation in these changes in coupling as a function of region and condition.

This pattern of results is somewhat challenging to explain using the theories that motivated these analyses. One approach is to revise the neural embellishment rather than the cognitive theory proper; for example, perhaps working memory updating can be accomplished by multiple neural substrates, including parietal cortex and the basal ganglia. But this seemingly innocuous revision then calls into question the necessity of a division of labor in the first place, given that other evidence suggests parietal cortex is apparently involved in maintenance operations as well (Postle, et al 2006; Edin et al., 2009). Perhaps these fundamentally different processes can take place within subregions of parietal cortex that are difficult to disentangle here. Alternatively, fundamentally different kinds of operations might take place in parietal cortex at different levels of recruitment. Yet another possibility is that the underlying cognitive theory is itself incomplete, because it fails to account for the interaction of updating with informational abstraction that the brain reflects. Along these lines, Nee & Brown suggest that alternative cognitive theories of the task, such as those which posit shifts of attention among items that have been updated into working memory, may be more naturally consistent with these results.

*Connectivity Analyses – Inferential Risks.* PPI is a complex technique that is, like many fMRI analyses, substantially underpowered – particularly for event-related designs. This fact reaffirms the necessity of comparing effects directly via formal tests of interaction, as opposed to (incorrectly) inferring that a significant result can be said to be different from a non-significant one (Niewenhuis, Forstmann & Wagenmakers, 2011). Such concerns about power are exacerbated by many early PPI analyses, which neglected to encode all conditions of the experimental design and thus may have further inflated the error term (McLaren et al., 2012).

More generally, the problem which PPI seeks to solve is fundamentally ill-posed. The inference we would like to make from a PPI analysis is that some task variable has induced a differential information flow between regions. But this requires testing whether BOLD in any region is correlated with the convolved interaction of neural activity in another region and the current task condition, not with the interaction of convolved neural activity and the convolved task condition. These two expressions are mathematically distinct (Figure 6A).

Unfortunately, to actually calculate the former requires blind deconvolution. That is one must work backwards from the BOLD timeseries to the underlying neural activity, a problem which can have non-unique solutions. To see why, imagine that you have observed a miniscule BOLD response at some particular time. Does this response reflect the peak of a similarly miniscule hemodynamic response at that same moment, or only the trailing edge of a much earlier and larger hemodynamic response? Even when information is pooled across an entire timeseries, these questions can have multiple answers, unless identifiable biophysical models are used (Friston, 2011, but see Lohmann, Erfurth, Müller & Turner, 2012; Lohmann, Stelzer, Neumann, Ay & Turner, 2013). Any errors introduced during deconvolution will also yield errors in the PPI regressor (see also Figure 6B), which will in turn reduce statistical power (via regression dilution; Frost & Thompson, 2000). Thus, while different fMRI analysis packages currently take different approaches on this point, the consequences for real PPI results are as yet unclear.

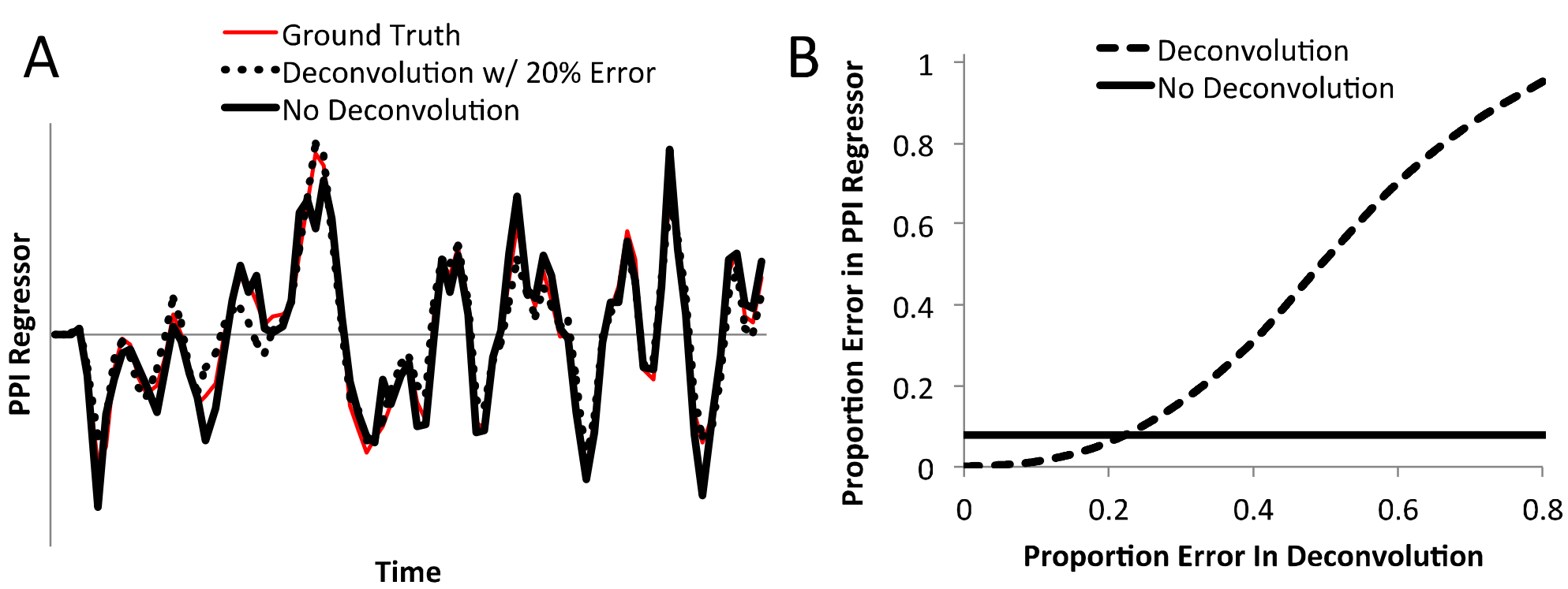

**Figure 6~~7~~. PPI requires solving the ill-posed problem of blind deconvolution, which could induce variable amounts of error; by contrast, well-posed alternatives are slightly though certifiably incorrect. A.** The mathematically-correct PPI regressor is the convolved product of an underlying neural timeseries with a task regressor ("ground truth"; red line). The underlying neural timeseries, however, is not observable using fMRI; only the BOLD response is. One approach to this problem (currently taken in the SPM package for fMRI analysis) is to calculate a proxy for the underlying neural timeseries via blind deconvolution. This proxy is then multiplied by the task regressor and convolved to yield predictions that may be close to ground truth (dotted line) – especially when the errors introduced by deconvolution are less than 20%. A qualitatively similar regressor is produced by avoiding blind deconvolution altogether, instead taking the product of the observed BOLD timeseries and the task regressor (solid line). This approach (currently taken by the FSL package for fMRI analysis) is mathematically incorrect, but nonetheless yields reasonably accurate results. Methods of these simulations are presented in [2]. **B.** The results of these approaches can differ wildly. When avoiding blind deconvolution, a small but constant error is introduced into the estimated regressor (solid line) due to the mathematical incorrectness of the procedure. By contrast, the consequences of blind deconvolution for the

---

[2] *Methodology of PPI Simulations: Figure A shows the first 100 timepoints of a 10,000 sample Monte Carlo simulation. This simulation assumes a canonical hemodynamic response function and i.i.d. Gaussian variables corresponding to task, neural, and error timeseries. Figure B shows summary statistics for multiple such simulations, each assuming various amounts of error in the deconvolution step along the x-axis. On the y-axis, proportion error is calculated as (1 –* $R^2$*), where R is the Pearson correlation between the ground truth regressor and the PPI regressor yielded by each approach. These errors are introduced either by the mathematical incorrectness of "No deconvolution" procedure itself (solid line), or by errors arising during blind deconvolution (dotted line). This latter source of error is also calculated as (1-*$R^2$*), where R is the Pearson correlation between the results of deconvolution and the underlying neural activity of each simulation.*

calculated regressor may be much larger or somewhat smaller (dotted line), depending on how much error this procedure introduces (which could vary across time, voxels, or subjects).

PPI is but one instance of a larger class of fMRI connectivity methods. Another instance of fMRI connectivity methods – indeed, a class unto itself – can be broadly described as variations on the theme of matrix factorization. In this case, the matrix is made up of 4 dimensions: one temporal, and three spatial. Well-known examples of this class include principal components analysis (PCA) and independent components analysis (ICA). Both ICA and PCA seek to identify separate underlying spatial patterns of activation ("components") that can be recombined (according to a time-varying weighting function) to yield the original data. ICA in particular has lead to a number of advancements in our understanding of the so-called "resting state" BOLD signal observed in the absence of any experimental task. Sophisticated refinements to these techniques allow the spatial factors to enjoy some degree of consistency across subjects, either by including subject as a separate dimension of the matrix (tensor ICA; Beckmann & Smith, 2005), by using clustering methods to identify how subject-specific ICA analyses correspond (Esposito et al 2005), or by regressing the time-varying weighting functions found at the group level onto the subject-specific datasets ("dual-regression"; Beckmann et al., 2009). Identifying the best methods for combining these matrix factorizations across subjects remains an important area for future work (e.g., Blumensath et al., 2013; Manning et al., 2014).

A rapidly-expanding family of matrix factorization methods imposes further constraints, in an attempt to improve interpretability. For example, topographic factor analysis (TFA; Manning et al., 2014) requires the spatial components to be independent spheres. Topographic latent source analysis (TLSA; Gershman et al 2014) goes further by requiring each of these spherical spatial components to have some degree of correspondence with the experimental design.

The aforementioned connectivity techniques can be understood as imposing (as in TFA and TLSA) or requiring (PPI, ICA and PCA, at least for statistical significance) some degree of spatial homogeneity in the components that are recovered. Accordingly, this broad class of connectivity methods may all share a number of inferential risks (e.g., O'Reilly et al., 2012; Cole, Smith & Beckmann, 2010). First, connectivity methods are even more sensitive to transient head motion and various cardiac and respiratory artifacts than standard univariate fMRI techniques (van Dijk, Sabuncu & Buckner, 2012; Power et al., 2012). This also means that the parametric predictors are measured with error, again raising the specter of regression dilution (Frost & Thompson, 2000). Finally, spurious alterations of regional connectivity can be provably induced by standard preprocessing steps, like global signal regression (Murphy, Birn, Handwerker, Jones & Bandettini, 2009; Saad et al., 2012), suggesting that care must be taken to ensure the robustness of any result achieved via these techniques.

Connectivity methods also typically involve implicit biophysical assumptions that may produce bias. As one example, the results of fMRI connectivity are potentially limited to a specific spectral range. Certainly connectivity changes taking the form of coordinated power fluctuations within the theta band (4-7Hz) could fail to be detected, given the putative absence of such theta effects in the BOLD response (de Munck et al., 2012). Of course, without an explicit biophysical model that specifies which frequency bands are responsible for changes in connectivity, and how these bands translate into BOLD, it is impossible to know how this insensitivity may distort our views of the functional connectome (see Hipp & Siegel, 2015, for recent evidence along these lines). As a second example, consider that connectivity and task-induced changes in connectivity are in principle wholly independent of mean effects on the BOLD response, but that many connectivity and PPI analyses nonetheless often rely on region definitions that are informed by (or even wholly based on) mean BOLD effects. Again, the true biophysical relationship between changes in mean BOLD and connectivity is unknown, but the implicit assumption of a relationship could certainly bias the inferences being drawn. As a final example, the dependent measure of some connectivity analyses involves the beta estimates resulting from regression of one timeseries on another. However, these betas reflect the *slope* relating BOLD across regions, not its *correlation*. Correlation and slope can be independent. The use of slope instead of correlation reflects another implicit assumption – namely, that slope is more biophysically indicative of connectivity – even though slope can be affected by manipulations that have no effect on the mutual information shared across two regions. For example, slope but not correlation will be necessarily affected by manipulations that shrink the range of one region's timeseries. Thus, numerous implicit biophysical assumptions could color the kinds of inferences currently being drawn on the basis of fMRI connectivity methods.

Beyond these issues in sensitivity, evidence of a functional network requires two features: the constituent regions must participate in a network, and the activity-dependence of this network on some function. Correlation among separate regions provides evidence for the first of these features. However, evidence for activity-dependence can come from either mean BOLD change, and/or changes in correlation (i.e., PPI), across conditions. It is often assumed that both mean BOLD change and PPI are necessary to assign function to a network. However, metabolic change associated with neural firing and correlations associated with neural synchrony are independent, and need not coincide.

*Connectivity Analyses – Summary*. Though not free of caveats, these connectivity methods represent a fundamentally novel (rather than merely "scaled up") class of analysis methods relative to those that are typical of behavioral analysis. These methods are likely to enable particularly direct mappings to the cognitive level of analysis, given that cognitive processes are most likely to be subserved by the action of multi-regional brain networks, rather than independent spatially-circumscribed regions. As with any method of such impressive flexibility and power, there are ways that connectivity analyses can go wrong. But this should not act as a deterrent from their widespread adoption, given their potential promise for illuminating the neural basis of cognition. And, as described below, these powerful and important connectivity methods have helped to enable other even more drastic departures from the familiar world of spatially-circumscribed univariate analyses.

*Multivariate Pattern Analysis (MVPA).*

In one particularly diverse class of methods, the spatial-contiguity constraints typically imposed on fMRI results (for controlling false positive rates) are relaxed, or even wholly eliminated. These more general methods allow researchers to test for "information" in the BOLD response, and to utilize cross-validation methods as an alternative means of controlling false positives. These techniques can detect information that more standard analyses cannot – that is, information other than spatially-contiguous differences in the raw BOLD response or its correlation across regions. For example, information might instead be contained in the *relative* patterns of BOLD or its correlations across voxels within some proximity (say, 8mm proximity, as in popular "searchlight" style analyses), within some region (e.g., the inferior frontal gyrus; as in ROI-based approaches) or even within the brain as a whole.

These approaches rely on rapidly developing machine learning techniques designed for high-dimensional problems. Disregarding even the (high) likelihood of future enhancements to these approaches, MVPA still promises to be of very widespread utility in cognitive neuroscience. This promise derives in part from the breadth of questions amenable to MVPA. For example, MVPA can help to identify the specificity of information being encoded in the brain (e.g., abstract vs. concrete; Wang, Baucom & Shinkareva, 2012); the precision of the encoding (e.g., Emrich, Riggali, LaRocque & Postle, 2013); the mutual information shared across these encodings (e.g., conjunctive vs. compositional; Cole, Etzel, Zacks, Schneider & Braver, 2011), and of course their spatial arrangement (sparse/distributed vs. dense/local; Hanson, Matsuka & Haxby, 2004). Many such representational issues have long been a core aspect of cognitive theorizing. MVPA therefore represents a major way that fMRI can be used to test (neurally-embellished) cognitive theories. Naselaris (this volume) offers a more technical primer on the use of these methods for such purposes.

As just one example of this utility, consider the long-standing hypothesis that one source of forgetting is proactive interference (PI), i.e., the blocking, suppression, or disruption of a to-be-remembered item by other information also in memory. Cognitive theories of PI vary, with some theorists proposing that PI serves to disrupt memory encoding (e.g., Wickens, 1970) and others proposing that it disrupts retrieval (e.g., Crowder, 1976). These debates may have remained unresolved in part because behavioral measures alone are insufficient for conclusively locating PI to the encoding or retrieval stage. For example, in the "release from PI" paradigm (Wickens, 1970), subjects are tested on a multiple lists of items; PI gradually builds when consecutive lists involve items of the same category, consequently decreasing memory performance. It is thought that such effects reflect reductions in the discriminability of items arising from the increasing activation of features shared across items in a category, at the expense of features that would distinguish each item from the others. Critically, however, such effects could occur either at encoding or retrieval (Öztekin & Badre, 2011).

The use of classifiers to analyze distributed patterns of BOLD – one form of MVPA – is an fMRI technique which is particularly suited to this problem. MVPA can allow one to quantify the degree to which categorical information is being represented (e.g., Haxby et al., 2001), even in the absence of any overt response. Öztekin & Badre (2011) employed this technique during a release from PI paradigm to show that neural representations of the category were incrementally *less* decodable across successive lists at the time of encoding – wholly contrary to the notion that increases in the presence of such category-general features might mediate PI during encoding. The opposite effect was observed during retrieval, consistent with the notion that PI is mediated by growing interference from category-general features specifically at the time of retrieval. Moreover, only these dynamics at the time of retrieval correlated with behavioral measures of PI; they did so with significantly greater strength than the analogous correlation during encoding.

This example clearly illustrates a case in which behavioral data alone was unable to resolve a key cognitive debate; by contrast, fMRI enabled inferences about a latent process that were otherwise unobservable. The requisite neural embellishments to cognitive theory were both minimal and straightforward: that category-general information content should be linearly decodable from the brain.

*Multivariate Pattern Analysis – Inferential Risks.* The example above illustrates only one form of MVPA; there are many (e.g., Norman et al., 2006; Pereira, Mitchell & Botvinick, 2009; Kriegeskorte & Kievit, 2013). As a relatively recent form of fMRI analysis, our understanding of its limitations and risks is still rapidly developing. Several key methodological and inferential variables have nonetheless been identified as key to the valid interpretation of MVPA results, and a more comprehensive introduction can be found elsewhere (Davis & Poldrack, 2013). We list just a few key issues here.

As in many other forms of fMRI analysis, there is a substantial risk for bias. In this case the risk arises because the techniques used for MVPA are typically highly parameterized, and these parameters can often require tuning. If tuning of a classifier's parameters occurs on the same data set that will be used for statistical inference, there is the risk that its accuracy of classification will be upwardly biased. This form of accidental look-ahead error is often known as "peeking."

Another potential source of bias is temporal autocorrelation. The BOLD responses to any given stimulus will tend to be more similar with those arising from any other stimulus when the stimuli occur in the same "run" of the scanner. Cross-run cross-validation methods are key for ensuring such temporal autocorrelation does not exert undue influence on classification accuracy or estimates of representational similarity across stimuli.

A third source of bias is the fact that the *direction* of differences between conditions in MVPA is commonly lost in typical analysis methods (Todd, Nystrom & Cohen, 2013). For example, if a classifier's accuracy in distinguishing two conditions is merely averaged across subjects, then one may achieve a group effect that shows a reliable difference between two conditions even if the direction of that difference is completely unsystematic across subjects.

This direction-insensitivity may render MVPA results more prone to subject-specific and other idiosyncratic confounds, quite distinct from the more traditional kinds of confounds that psychologists are accustomed to eliminating. Consider a situation where variations in reaction time across two conditions are of sole importance, and accuracy differences are considered nuisance variance. Statistical inference will be safe at the group level, so long the conditions are compellingly matched in aggregate accuracy. However, if scanning with fMRI also took place, a significant group-level MVPA result could easily obtain for unwanted reasons. In this case, subject-specific classifiers could simply learn to recognize the subject-specific direction of differences in accuracy that characterize conditions. If the performance of these classifiers was then averaged to produce a group-level summary statistic, this group-level average would simply show above-chance classification. Such a result could easily, though incorrectly, be taken as a difference not attributable to accuracy, since accuracy was matched across conditions.

*MVPA – Summary.* MVPA is typically used to address two types of questions in testing a neurally-embellished cognitive theory. First, it can be used to ask what type of information is represented by a particular part of the brain. To address this type of question, one must rule out all possible variables correlated with the variable of interest (i.e., proxies). Though an issue with any type of experimental design (e.g., functional localization), issues unique to MVPA like directionless statistics make this particular problem daunting for decoding. The second type of question asks what information is available to the system across conditions or stages of a process. In these cases, whether one is decoding the representation of interest itself or a proxy for that representation may not bear on the inference. However, this logic rests on the assumption that the relationship between proxies and the information of interest holds across the training and test conditions. This assumption can be difficult to support without knowing the nature of the proxies, or the mechanism making them diagnostic of the underlying representation of interest. Despite these limitations, MVPA is amenable for use in assessing neural embellishments that lie close to the core of cognitive theorizing, including issues related to representational specificity, precision, compositionality and format.

*Topological Structure.*

It has been claimed that fMRI cannot help to choose among cognitive theories unless each theory specifies “the necessary spatial arrangement of its components (rather than their topology)” (Page, 2006). However, new techniques involving the use of graph theoretic measures *do* permit investigations of pure topology in a way that is almost completely independent of their spatial arrangement. This, in turn, opens the door for fMRI to inform a class of extremely abstract cognitive theories for which any more concrete neural embellishment may currently seem too speculative.

One useful example is the case of the Global Workspace (GW) Theory of consciousness (Baars, 2005). The core claim of this theory amounts to the highly abstract notion that consciousness (however that is defined) is enabled by an array of cognitive processes that are topologically arranged so as to rapidly distribute information about the sensory world. Such a theory might seem untestable and abstract. Yet recent fMRI evidence reveals a picture of the conscious brain that is surprisingly and perhaps uniquely consistent with GW theory.

Achard et al. (2012) analyzed differences in the patterns of functional connectivity in comatose patients and healthy controls, using a series of topological measures borrowed from network science (for an introduction see Baronchelli, Ferrer-i-Cancho, Pastor-Satorras, Chater & Christiansen, 2013). Their topological approach entailed numerous analyses of the structure of correlations across more than 400 different neural regions – including metrics like the mean correlation, probability of a connection between two regions that both connect with a third (i.e., the clustering coefficient), minimum number of regions that must be traversed to connect any two regions (i.e., the average path length), and many others. Of these myriad measures, the few to reliably distinguish the comatose from healthy controls all concerned the relative degree of centrality, or "hubness", of these regions.

Critically for GW theory, there was a robust negative correlation between the regions identified as hubs in the healthy controls and those identified as hubs in the comatose. Such a result is tantalizingly consistent with the most naïve neural embellishment of GW theory, in that like the proposed cognitive topology, network topology might also be closely associated with consciousness. More impressively, most of the regions showing a down-regulation in hubness in the comatose were located in primary visual cortex. This finding, too, is strikingly consistent with the naïve embellishment of GW theory that the neural topology of consciousness is specifically relevant to sensory information.

Though this evidence does not conclusively support GW theory, the evidence should nonetheless increase our confidence in it. And, more to the present purposes, fMRI provided this evidence in a domain where it would be impossible to collect behavioral responses whatsoever, in that the key test involves subjects who are unresponsive by definition. Finally, we note again that the requisite neural embellishment was both minimal and straightforward: one needed to assume merely that the topological structure of the cognitive operations posited by GW theory should be visible in the topological structure of the BOLD response at rest.

*Topological Structure – Inferential Risks.* Like MVPA, topological structure of functional connectivity is a rapidly-developing field, and our understanding of its inferential risks are evolving in tandem. Care must clearly be taken in the underlying measures used to calculate connectivity, and the benchmarks with which these measures are compared; for example, linear correlation may in some cases produce a bias towards the appearance of "small worldness" – a topological feature of much theoretical interest in the study of natural networks (Zalesky, Fornito & Bullmore, 2012).

Many topological investigations also pose a tricky problem at the outset: how finely should we discriminate neural regions, and which do we discriminate in the first place? It seems likely that topological metrics are sensitive to such choices. In the worst-case scenario, this could mean that the neural embellishment necessary to conduct a topological test of cognition would implicitly require a theory of the functional organization of the entire brain. Fortunately, in practice, major topological features do not seem particularly sensitive to these choices, at least when the number of regions analyzed is sufficiently high (e.g., Craddock, James, Holtzheimer, Hu & Mayberg, 2012).

Finally, we note that replacing regions with networks does not by itself alleviate any of the inferential problems that can affect other fMRI approaches, as described above. Simply exchanging voxels for networks does not fundamentally eliminate concerns about how to draw valid inferences from double dissociations or parametric effects, or how to address statistical concerns related to power and bias.

*Topological Structure – Summary.* Topological measures derived from graph theory are another relatively recent addition to the set of fMRI analysis tools. Like MVPA, these tools offer unique advantages for testing neurally-embellished facets of abstract cognitive theories. For example, these new tools make it possible to assess whether measures of the connectedness of a hypothesized cognitive system corresponds with identical measurements of the neural system, without requiring one to specify the spatial location or arrangement of the cognitive system's parts within the brain. There are inferential dangers associated with the use of these tools – including the counterintuitive risks associated with the use of fairly standard statistics, like linear correlation. Nonetheless, this class of analysis methods further extends the role of fMRI in testing cognitive theory.

**Working from the Bottom Up: Testing Emergentist Theories**

We have so far largely focused on cases where a theory was generated in the behavioral or purely cognitive domain, and then is embellished so as to be tested using neuroscientific data. However, there are a number of models of cognition that were initially developed on the basis of neuroscientific, rather than behavioral or cognitive data. These models range from basic neural networks to elaborated models of multiple interacting brain systems. Using this more bottom-up approach, the theorist attempts to apply known neuroscientific features to the model in an effort to explain or make predictions in the emergent cognitive and behavioral domain. Examples of this include the use of a unified set of biologically-inspired learning rules and membrane potential dynamics to explain a variety of perceptual, semantic, grammatical, attentional and developmental phenomena (e.g., O'Reilly et al. 2012).

One would imagine that such models would be readily amenable to testing with fMRI. However, it has been strikingly difficult to find strong tests of these and similar models in the imaging literature. Indeed, far from the compelling parametric designs that test quantitative predictions generated by abstract mathematical models, as discussed earlier (e.g., prediction error), tests of these neurally-realized models using fMRI have primarily been qualitative and indirect.

As one (counter) example, Badre and Frank (2012), utilized a neural model of the corticostriatal system to predict fMRI data from a hierarchical control task. In this task, subjects learned via trial and error to press one of three buttons corresponding to stimuli varying in color, shape, and orientation. Unbeknownst to subjects, there were two conditions. In the hierarchical condition, one stimulus dimension deterministically specified which of the others would be relevant to the response. In the flat condition, no such rule could apply; instead, the mappings of stimulus features to responses were conjunctive and requiring simple memorization.

Bottom-up models of the human corticostriatal system imply that learning in the hierarchical, but not flat condition should rely crucially on the reinforcement learning signals taking place in a more rostral corticostriatal loop (Frank & Badre, 2012). However, such models are incapable of generating unique trial-by-trial predictions for test with fMRI because of the large number of parameters. Instead, an intermediate algorithmic model was constructed – in this case, a Bayesian "mixture of experts" – that could be used to identifiably link the bottom-up neural network model with the trial-by-trial choices made by human subjects in the scanner.

Why should there be so few examples in the literature of testing these bottom-up theories with fMRI? One possibility is that the list of potentially-relevant biophysical parameters is generally so large that behavioral simulations must cull the list aggressively, so as to remain identifiable and computable in reasonable time. In this process, parameters which render BOLD identifiable may be incidentally culled or combined. One example of this is the winner-take-all approximation to the influence of locally-connected inhibitory interneurons (e.g., O'Reilly et al, 2012). This shortcut is useful for reducing the parameter space in behavioral simulations but, at the same time, complicates the development of straightforward predictions for BOLD, which is thought to be differentially sensitive to gradient restoration in inhibitory interneurons (Buzsáki, Kaila & Raichle, 2007).

Perhaps counterintuitively, more abstract reinforcement learning and Bayesian models might actually have more direct mappings to BOLD. Both reinforcement learning and Bayesian models tend to foreground the updating of information content – whether via Bayesian update, or the Rescorla-Wagner delta rule. Such information content changes may be in turn quite directly linked to energetic demands. Numerous theories propose that changes in information content necessitate the movement of a neural population from one local minimum to another within an energetic landscape (e.g., Friston, 2010). If these minima are to be locally stable, then transitioning among them must require some degree of "hill-climbing" in this landscape. Put another way, changes in information content require net energetic input. Thus, models that focus on information updating may have advantages in explaining BOLD relative to models which go on to elaborate only the partial subset of the underlying biophysics necessary for behavior (but perhaps not BOLD).

Nevertheless, these more elaborated neural theories do have some clear advantages. First, they seem to have proven comparatively more successful than their abstract counterparts in predicting the relationship of cognition to other kinds of neuroscientific data (including the cognitive effects of drugs and disease; e.g., Frank, Seeberger & O'Reilly, 2004). Second, they may be more amenable than their abstract counterparts to tests with newer fMRI techniques, including analyses of connectivity (Horwitz et al., 2005), topology (Eldar, Cohen & Niv, 2013) and other multivariate characteristics (Kriegeskorte & Kievit, 2013).

To summarize, fMRI offers opportunities for testing not only abstract cognitive theories, but also those that have been built "bottom-up" – in other words, those given a neural implementation in formal mathematical models. Many such theories are rooted in microscopic neuronal dynamics and the learning rules that govern their alterations with experience, and might therefore be expected to have a particularly close relationship with biophysical measures like BOLD. This makes the lack of contact between fMRI and these emergentist models all the more surprising. One reason for this lack of contact may be that energetically-important parameters are often simplified in emergentist models, to enable efficient simulations of behavior. Counter-intuitively, models with more abstract implementations have faired somewhat better, perhaps because they tend to focus on energetically-demanding updating processes (whether of action values or of posterior distributions, in reinforcement learning and Bayesian models respectively). Nonetheless, we suggest that emergentist models may be more easily aligned with pharmacological, connectivity, and multivariate fMRI techniques than more abstract cognitive theories and associated mathematical models.

## In Defense of Localization and Reverse Inference

Above we've provided several key examples of how fMRI has informed cognitive theory in situations where the key behavioral data was either intrinsically ambiguous or impossible to acquire in the first place. In every case some form of neural embellishment is required. In most cases, the assumptions involved in the requisite neural embellishments were natural and straightforward. Of course, such straightforward neural embellishments will not always be right. Ideally, we would have a principled basis for even more specific kinds of neural embellishments. Basic work in localizing cognitive processes is important in this regard – even when such work does not directly inform *any* cognitive theory. Thus, in answer to Fodor's question, quite a bit may turn on "precisely how far north of the neck" the neural correlates of hypothesized cognitive processes are. We believe this to be the case for two reasons.

First, pure localization work is critically important for enabling other kinds of tests of neurally-embellished cognitive theory – including causal tests like those enabled by transcranial magnetic stimulation, and tests of high-resolution temporal dynamics like those enabled by EEG. Many of the limitations of fMRI we described above do not pertain to these other technologies, a fact which naturally motivates an iterative, convergent approach.

Second, pure localization work may be key for a more principled neural embellishment of cognitive theory – as well as for the development of future neurally-*inspired* theories such as those discussed above. We understand a great deal about the brain. And, one fact is that the brain is not homogenous. Rather, individual regions of the brain have features relevant to structure, neurochemistry, and interactions with other regions that will impact the type of information processing that they do. Localization can provide important information about the involvement of these specific sites during cognition. This information is particularly important for building neurally elaborated models that predict emergent system behavior.

There are other reasons that the identity of the neural structures engaged by a task could be informative, even in the absence of such a model. For example, when performed with the appropriate statistical inference (e.g., Bayesian), one *can* probabilistically infer the relation of one task (or cognitive construct) to others given activity in a particular part of the brain (Poldrack, 2011; Yarkoni, Poldrack, Nichols, Van Essen & Wager, 2011; Machery, in press). Such approaches demonstrably work, at least in the visual domain (Kay, Naselaris, Prenger & Gallant, 2008; Nishimoto et al., 2011), and projects currently underway seek to extend this approach to other domains (including, for example, Neurosynth: http://www.neurosynth.org). If this optimistic goal can be accomplished – by no means a foregone conclusion – the cognitive dynamics of naturalistic behavior might some day be probabilistically inferred purely from an fMRI scan. The potential for such technology in testing cognitive theory is hard to overestimate.

In summary, the science of localizing cognitive functions to neural substrates is sometimes derided as irrelevant – as in Fodor's famously pithy quote: "If the mind happens in space at all, it happens somewhere north of the neck. What exactly turns on knowing how far north?" (Fodor, 1999). We disagree: the neural embellishments required for neuroscientific tests of cognitive theory are strongly informed by these localization studies. In addition, localization studies may ultimately form the vocabulary used by machine learning algorithms to identify the cognitive underpinnings of highly complex or even naturalistic behaviors.

**Conclusion**

We have argued that there are many advantages to neurally embellishing a cognitive theory. Most plainly, doing so makes a theory more general and falsifiable and more robust to the risks of model mimicry and descriptive abstractions. Neural embellishments, like behavioral embellishments commonly known as operational definitions, involve specifying a spatiotemporal functional form; this involves substantial assumptions. We have described how these embellishments nonetheless enable fMRI to inform diverse cognitive theories in domains like planning, learning, working memory, long-term memory, and even consciousness. For each example, we have also noted the many inferential risks that accompany these data. Such limitations are not unique to the use of fMRI, but rather advocate an integrative, iterative approach to testing neurally-embellished cognitive theory. This approach will necessarily rely in part even on the most humdrum "localization" aspects of fMRI research, which can ultimately also be informative for cognitive theory, albeit often in empowering other more direct experiments. In conclusion, we think that neural embellishment generally, and more specifically the type that is testable using fMRI, can be valuable to cognitive scientists, even those whose interests do not primarily involve the brain.